\documentclass[aps,pre,
 11pt, 
   preprint,
    tightenlines,
   superscriptaddress,
   eqsecnum,
   floatfix,
   showpacs,
   preprintnumbers,
   longbibliography,
   nofootinbib
   ]{revtex4-1}
\usepackage{color}
\usepackage{graphicx}
\usepackage{framed}
\usepackage{subfigure}
\usepackage{stackrel}
\usepackage{amsmath}
\usepackage{amssymb}
\usepackage{time}
\usepackage{graphicx}
\usepackage{enumerate}
\usepackage{latexsym}
\usepackage{bm}
\usepackage{upgreek}
\usepackage{mathrsfs}
\DeclareSymbolFont{AMSb}{U}{msb}{m}{n}
\DeclareSymbolFontAlphabet{\mathbb}{AMSb}
\usepackage{physics}

\begin{document} 
\newcommand{\bq}{\begin{equation}}
\newcommand{\be}{\begin{equation}}
\newcommand{\ba}{\begin{eqnarray}}
\newcommand{\eq}{\end{equation}}
\newcommand{\ee}{\end{equation}}
\newcommand{\ea}{\end{eqnarray}}

\date{\today, \now \ MDT}
\title{A Generalized Nonlinear extension of Quantum Mechanics}

\author{Alan Chodos}
\email{alan.chodos@uta.edu}
\affiliation{Dept. of Physics, University of Texas at Arlington, 502 Yates Street, Box 19059, Arlington, TX 76019 }

\author{Fred Cooper} 
\email{cooper@santafe.edu}
\affiliation{The Santa Fe Institute, 1399 Hyde Park Road, Santa Fe, NM 87501, USA}
\affiliation{Theoretical Division and Center for Nonlinear Studies,
   Los Alamos National Laboratory,
   Los Alamos, NM 87545}
\date{\today}
\begin{abstract}
We construct the most general form of our previously proposed nonlinear extension of quantum mechanics that possesses three basic properties. Unlike the simpler model, the new version is not completely integrable, but it has an underlying Hamiltonian structure.  We analyze a particular solution in detail, and we use a natural extension of the Born rule to compute particle trajectories. We find that closed particle orbits are possible.
\end{abstract} 
\maketitle

\vspace{1\baselineskip}
\section{Introduction of the model}



In a previous paper \cite{cf23} , we proposed a non-linear extension of quantum
mechanics and studied some of its consequences. Our motivations were
various: partly, in the spirit of \cite{Birula},  \cite{Weinberg89} and  \cite{Kaplan}, we simply wanted to
see what a particular non-linear extension of quantum mechanics might
look like, and to deduce some of its consequences.

But we also had in mind the question of whether, as others have
speculated \cite{Penrose} \cite{Carlip} \cite{Pang}, the union of gravity with quantum mechanics is best
achieved not by producing a fully quantum version of general relativity,
as is contemplated for example in string theory or loop quantum gravity,
but by some other means, such as the coupling of quantum mechanics to a
classical version of gravity, as in the work of Oppenheim and
collaborators \cite{Oppenheim} , or indeed by the modification of quantum mechanics
itself.

In the model we introduced in  \cite{cf23, cf24}, a given system is described by 2
state vectors, \(|\psi >\) and \(|\phi >\), in Hilbert space, in
contrast to the single state vector to which ordinary quantum mechanics
limits itself.  Our procedure is heuristic, that is, we do not impose a prior rationale for introducing two state vectors but instead look ahead to what the consequences are. In that sense, we follow the grand tradition of quantum mechanics, in which everybody knows how to calculate, but nobody knows what it all really means. Our approach affords us the opportunity to generalize quantum mechanics in what we consider to be an interesting way. In our first paper, we presented the minimal realization of this idea, which, surprisingly, allowed us to completely determine the effects of the non-linearity independently of the dynamics of the underlying quantum mechanical system. This simplicity, however, overly restricted the evolution possibilities.
The nonlinearity in that model is proportional to the inner product $\gamma = < \phi | \psi>$. One finds that $\gamma \rightarrow  0$ for large times, so that the effect of the nonlinearity vanishes in the far future and the far past. 

 In this paper, we retain the same principles but allow for a greater range of dynamical effects at the cost of no longer being able to solve the system of equations exactly. 
Rather than speculate about the meaning of the second state vector, we study the effects it produces, thereby hoping to gain information on what meaning it might have. One line of investigation, which we began to consider in \cite{cf24} , is that the additional dynamics have something to do with gravity. In other words, perhaps gravity is, at least in part, a manifestation of the non-linear aspects of quantum mechanics that we have introduced. The fact that our non-linear modification is universal, i.e. independent of the Hamiltonian of the underlying system, may be relevant in this regard. In section III we define a trajectory function, which describes how a particle would move under the influence of the non-linear terms, and compute it in a simple case.


If we define

\begin{equation} 
\vert \Psi > = \begin{pmatrix}
\vert \psi > \\ \vert \phi > \\
\end{pmatrix} 
\end{equation}

and

\bq~~
\mathcal{H =}\begin{pmatrix}
H & 0 \\
0 & H \\
\end{pmatrix}
\eq
where $H$ is the Hamiltonian of the underlying system, then the
equations of motion of the model in reference \cite{cf23}  are

\bq~~ i\frac{\partial|\Psi >}{\partial t}\mathcal{= H|}\Psi\mathcal{> + M|}\Psi >\eq
where \(\mathcal{M}\) is the Hermitian matrix

\bq~~\mathcal{M =}\begin{pmatrix}
0 & g < \phi|\psi > \\
g^{*} < \psi|\phi > & 0 \\
\end{pmatrix}. \eq
The coupling constant \(g = a + ib\) is in general complex
(\(b \neq 0)\) .

    We want to replace this \(\mathcal{M\ }\)with a more general one,
subject to the following conditions, all of which are obeyed by our
original model:

\begin{enumerate}
\def\labelenumi{(\Roman{enumi})}
\item
  Each element of \(\mathcal{M}\) should depend linearly on the inner
  products \(< \psi|\psi >\), \textless{}\(\phi|\psi >\),
  \(< \psi|\phi >\), and \(< \phi|\phi >\), and should depend on time
  only through them;
\item
  The equations of motion should imply that
  \(< \psi|\psi > + < \phi|\phi > = N = const.;\)
\item
  The equations of motion should be invariant under the separate
  constant phase transformations
  \(\left| \psi > \rightarrow e^{i\theta_{1}} \right|\psi >\) and
  \(\left| \phi > \rightarrow e^{i\theta_{2}} \right|\phi >\) .
\end{enumerate}

The reason for condition (I) is that it encapsulates the philosophy behind our non-linear extension of quantum mechanics. There are infinitely many ways to try to extend quantum mechanics; the procedure we have chosen is to introduce a second state vector, and then to have the relevant inner products govern the dynamics of the non-linear extension. That is the content of condition (I).
Condition (II) is important for two reasons: it is the appropriate generalization of the usual rule that the norm of the state vector remains constant, and it ensures that the evolution of the state vectors will remain bounded despite the presence of complex coupling coefficients. Condition (III)  is a consequence of Occam's  razor. It is a symmetry that significantly restricts the number of allowed terms in the equations of motion.

To implement condition (II), it is sufficient for \(\mathcal{M}\) to be
Hermitian, but it is not necessary. \(\mathcal{M}\) can have an
anti-Hermitian part of the form

\bq~~\mathcal{M}^{'} = \begin{pmatrix}
i\mu < \phi|\phi > & \lambda < \phi|\psi > \\
 - \lambda < \psi|\phi > & - i\mu < \psi|\psi > \\
\end{pmatrix},
\eq
where \(\mu\) and \(\lambda\) are real constants. Then the most general
\(\mathcal{M}\) satisfying (I), (II), and (III) is

\bq~~\mathcal{M =}\mathcal{M}_{0}\mathcal{+ M'}\eq
where

\bq~~\mathcal{M}_{0} = \mathcal{M}_{0}^{\dagger} = \begin{pmatrix}
\alpha_{1} < \psi|\psi > + \alpha_{2} < \phi|\phi > & g < \phi|\psi > \\
g^{*} < \psi|\phi > & \beta_{1} < \psi|\psi > + \beta_{2} < \phi|\phi > \\
\end{pmatrix}.\eq

Here the \(\alpha_{i}\) and \(\beta_{i}\) are real constants, and
\(g = a + ib\) was defined above. In total, there are 8 real parameters,
as compared to two in the original model.

\subsection{The  $\tau$- $\delta$ problem}

If \(\mathcal{M} = 0\), all the inner products would be constant. Hence,
the inner products obey evolution equations that are independent of
\(H.\) In these equations, we find that most of the above parameters
play no role. With \(\gamma = < \phi|\psi > ,\ \) we define
\(\delta = |\gamma|^{2}\), and let

\bq \tau = < \psi|\psi > - < \phi|\phi >. \eq

The following differential equations for $\tau$ and $\delta$  are a direct consequence of the equations of motion, Eq. (1.3):
\ba  \label{tdeqs} 
\frac{d\tau}{dt}&& = \mu\left( N^{2} - \tau^{2} \right) + 4b\delta,  \nonumber \\
\frac{d\delta}{dt}&& = - 2(b + \mu)\tau\delta. 
\ea

The analogous equations for the original model are recovered for
\(\mu = 0\). To gain some preliminary insight, we note that we can
easily integrate eqs.(\ref{tdeqs})  in 3 special cases: (1) \(\mu = 0\); ~~(2)
\(b + \mu = 0;\ \) ~~(3) \(b = 0\) .

Case (1) is just the original model, for which we have that $\tau(t)$ and $\delta(t)$ are given by:
\bq
\tau(t)  = 2\omega_{0}\tanh{2\omega_{0}bt} ,~~
\delta(t)  = \frac{\omega_{0}^{2}}{\cosh^{2}2\omega_{0}bt} ,
\eq
where \(\omega_{0}\) is a constant of integration. In general, we expect
the solution of these 2 equations to depend on one constant of
integration in addition to the freedom to redefine the origin of time,
\(t \rightarrow t - t_{0},\) which we do not explicitly indicate.

Case (2): we find \(\tau = - 2\omega_{0}\tanh{2\omega_{0}bt}\) and
\(\delta = \frac{N^{2}}{4} - \omega_{0}^{2}\).

Case (3): we find \(\tau = N\tanh{\mu Nt}\) and
\(\delta = \frac{\delta_{0}}{\cosh^{2}\mu Nt}\) . In this case, the
constant of integration does not appear in the argument of the tanh
function, but only in the normalization of \(\delta\).

Noting that in all three special cases \(\tau\) is a hyperbolic tangent,
we can make the ansatz

\bq~~\tau = \alpha'\tanh{\beta't},
\eq
where \(\alpha'\) and \(\beta'\ \)are constants to be determined, and
plug this into the full equations for \(\tau\) and \(\delta\). We indeed
find a solution, namely

\bq~~\tau = N\tanh{(b + \mu)Nt}\eq
and
\bq~~\delta = \frac{N^{2}}{4}\left\lbrack \cosh{(b + \mu)Nt} \right\rbrack^{- 2} .
\eq

But this is definitely not the general solution, because \(\alpha'\) and
\(\beta'\) are fixed by the equations, so there is no free constant of
integration. Indeed, if we go through the special cases that we solved
above, we find that our solution to the general equations corresponds to
a specific choice for the constant of integration: Case (1),
\(\omega_{0} = \frac{N}{2}\); Case (2), \(\omega_{0} = 0\); Case (3),
\(\delta_{0} = \frac{N^{2}}{4}.\) 

\section{Hamiltonian formulation of the $\tau$ -$\delta$  system of equations}
In this section, we want to show that the differential  equations for $\tau$ and  $ \kappa= \ln \delta$ can be cast into a Hamiltonian framework, so that
a conserved energy can be identified. In terms of  the variables
 \(\kappa = \ln{\delta,\ }\) and   \(\Omega = \tau^{2}\), we have

\bq~~
\frac{d\kappa}{dt} = - 2(b + \mu)\tau
\eq
and therefore

\bq~~ \label{dtaudt} 
\frac{d\tau}{dt} =\frac{d\tau}{d\kappa}  \frac{d\kappa}{dt} = - (b + \mu)\frac{d\Omega}{d\kappa} = \mu\left( N^{2} - \Omega \right) + 4be^{\kappa}.
\eq

This is a first-order linear differential equation for \(\Omega\) as a
function of \(\kappa\), and can be integrated exactly:
\bq~~
\Omega = \tau^{2} = N^{2} - 4 e^{\kappa} - c e^{\frac{\mu\kappa}{b + \mu}} = N^{2} - 4\delta - c\delta^{\frac{\mu}{b + \mu}} .
\eq

Here \(c\) is a constant of integration. In the original model,
\(\mu = 0,\) and \(c = N^{2} - 4\omega_{0}^{2}\) . The relation above is
the generalization of \(\omega_{0}^{2} = \frac{\tau^{2}}{4} + \delta\)
(eq. (2.15) of \cite{cf23}) in the original model. Since

\bq~~ N^{2} - \tau^{2} - 4\delta = 4\mathcal{S} ,\eq
where
\bq~~
\mathcal{S}   =< \psi|\psi >   < \phi|\phi > - < \psi|\phi > < \phi|\psi >
\eq
is the Schwarz parameter, the relation can also be expressed as
\be
4\mathcal{S} = c\delta^{\frac{\mu}{b + \mu}} .\nonumber
\ee 
In the original model,
$\mathcal{S}$ is a constant, but in this more general case it
apparently will not be.

To solve these equations completely, we would need to substitute the
relation between \(\tau\) and \(\delta\) back into the equations for
\(\frac{d\tau}{dt}\) and \(\frac{d\delta}{dt}\) and perform one more
integration.

To simplify the subsequent discussion, we adopt a slightly different
notation. As evolution parameter, we choose \(s = - (b + \mu)t\), and
define \(p = \frac{\mu}{(b + \mu)}\).{[}Note: this requires
\(b + \mu \neq 0,\ \)so Case 2 above is not covered by this analysis.{]}
The equations of motion for \(\tau\) and \(\kappa\) are then

\bq~~
\frac{d\kappa}{ds} = 2\tau
\eq

and

\bq~~
\frac{d\tau}{ds} = - 4e^{\kappa} - pce^{p\kappa},
\eq
with the additional relation
\bq
N^{2} = h \equiv \tau^{2} + 4e^{\kappa} + ce^{p\kappa}  .
\eq
 We recognize
that \(h\) serves as the Hamiltonian for this system, with \(\kappa\)
playing the role of the coordinate and \(\tau\) the role of the
momentum. In fact the first order equations for $\kappa$ and $\tau$ are just Hamilton's equations:

\bq~~\frac{d\kappa}{ds} = \frac{\partial h}{\partial\tau}\eq

and

\bq~~\frac{d\tau}{ds} = - \frac{\partial h}{\partial\kappa}\eq

So \(\kappa\) behaves like the position $x$  of  a particle in 1-D, moving in the potential
\(V(\kappa) = 4e^{\kappa} + ce^{p\kappa}\). This explains in a more
general way why in the original model
\(\delta = |\gamma|^{2} \rightarrow 0\) for large times. In the original
model, \(p = 0,\) so \(V(\kappa)\) tends to infinity for large positive
\(\kappa\), but for large negative \(\kappa\) the potential tends
smoothly to a constant. Therefore \(\kappa\) escapes to minus infinity,
and \(\delta = e^{\kappa}\) tends to zero.

If \(p > 0\ ,\ \ \ V(\kappa) = 4e^{\kappa} + ce^{p\kappa}\) will have
the same general behavior as in the \(p = 0\) case. (The parameter \(c\)
is constrained to be positive because of the relation
\(ce^{p\kappa} = 4\mathcal{S).}\) But if \(p < 0\), \(V(\kappa)\) will
tend to infinity both for large positive and negative \(\kappa\),
insuring that \(\delta\) stays bounded away from either zero or infinity
as time evolves. In terms of the original parameters, assuming
\(b > 0\), this means \(- b < \mu < 0\) . This is the range of
parameters we shall consider going forward.

It is not known to us whether the problem of a classical 1-D particle
moving in a potential \(V(\kappa) = 4e^{\kappa} + ce^{p\kappa}\) with
\(c > 0\ \)and \(p < 0\) can be solved exactly. Below we present the
general solution for a particular choice of parameters, and another
particular solution that is valid for all choices of the parameters.

Energy conservation  leads to 
\be \label{ec} 
E=N^2 = \tau^2 + 4 e^\kappa+ ce^{p \kappa}, 
\ee
where from Hamilton's equations:
\be
 2 \tau = \frac{d \kappa} {ds} .
 \eq
Thus
  \be
  N^2 = \frac{1}{4} (\frac{d \kappa} {ds} )^2 + V(\kappa)
  \ee
   and 
  \be  \label{tr} 
  ds = \pm \frac{d\kappa}{2 \sqrt{N^2-V(\kappa)}}.
  \ee

\subsection{The case $p = - 1$} 

If we choose \(\mu = - \frac{b}{2}\), then \(p = - 1\).  
This also leads to $b+\mu=\frac{b}{2}$. so that $s=-\frac{b}{2} t $ Furthermore, we
can translate the coordinate: $\kappa = \eta + \eta_{0}$, where
\(\eta_{0} = \frac{1}{2}\ln\frac{c}{4}\) , so that the potential becomes
\(V(\eta) = 4{\sqrt{c}\ \cosh}\eta\).

The potential problem we want to solve is 
\bq~~
\frac{d^2 \eta}{ds^2}  = -\frac{dV}{d \eta } = - 4 \sqrt{c} \sinh(\eta). 
\eq
The solution is
\bq~~ \label{tr2} 
\eta = \pm 2 i ~ \text{am}\left(\frac{1}{2} \sqrt{\left(8 \sqrt{c}-c_1\right) (s+c_2){}^2}|-\frac{16
   \sqrt{c}}{c_1-8 \sqrt{c}}\right)
   \eq
   Here am $(u |m)$  is the Jacobi amplitude which is the inverse of the incomplete elliptic integral of the first kind  $F(u |m)$.  
   Note that to obtain this symmetric potential we needed to choose $\mu=-b/2$  so that the parameter $s= -\frac{b t}{2}$  where $t$ is the actual time.
 
\begin{figure}
\includegraphics[width=0.4\linewidth]{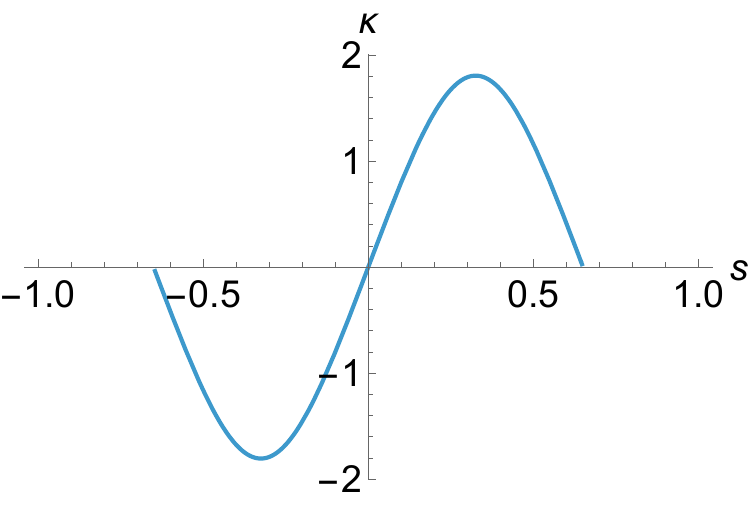}
\caption{  $\kappa (s)$ for $N^2=25,c=4$ }
	\label{kappa}
\end{figure}

Utilizing Energy conservation Eq. (\ref{ec}) we determine  these constants.  In particular  one has for the $+$ sign in Eq. (\ref{tr}) 
  \bq~~
  s=- \frac{i ~F\left(\frac{i \eta}{2}|\frac{8 \sqrt{c}}{4 \sqrt{c}-N^2}\right)}{\sqrt{N^2-4 .
   \sqrt{c}}}
   \eq
  Inverting this equation one obtains:
  \bq~~
\eta= -2 i~ \text{am}\left(i \sqrt{N^2-4 \sqrt{c}} s|\frac{8 \sqrt{c}}{4 \sqrt{c}-N^2}\right)
  \eq
  When $c=4, \eta_0=0$. so that $\kappa=\eta$.
We plot $\kappa(s)$  in Fig.( \ref{kappa}) for $c=4, N^2=25$ 

%
%
From  $\kappa $ we can determine  $\tau$
\bq~~ \label {taussym} 
 \tau(s) = \frac{1}{2} \frac{d\kappa}{ds} =\sqrt{N^2-4 \sqrt{c}} ~ \text{dn}\left(i \sqrt{N^2-4 \sqrt{c}} s|\frac{8 \sqrt{c}}{4
   \sqrt{c}-N^2}\right)
 \eq
where $~\text{dn} (u |m) $ is the delta amplitude, satisfying the relationship
\bq
    ~\text{dn} (u |m) = \frac{d}{du} \text{am}  (u |m). \nonumber 
  \eq.

\begin{figure}
\includegraphics[width=0.4\linewidth]{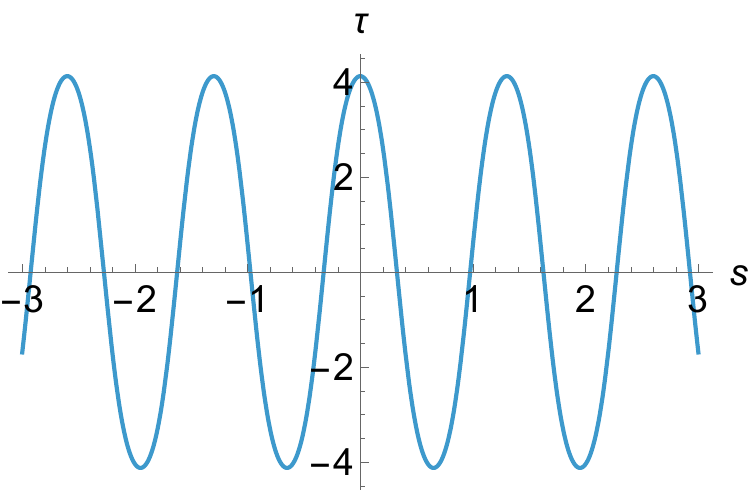}
\caption{  $\tau (s)$  for $N^2=25,c=4$ }
	\label{taus}
\end{figure}

\begin{figure} \label{deltas}
\includegraphics[width=0.4\linewidth]{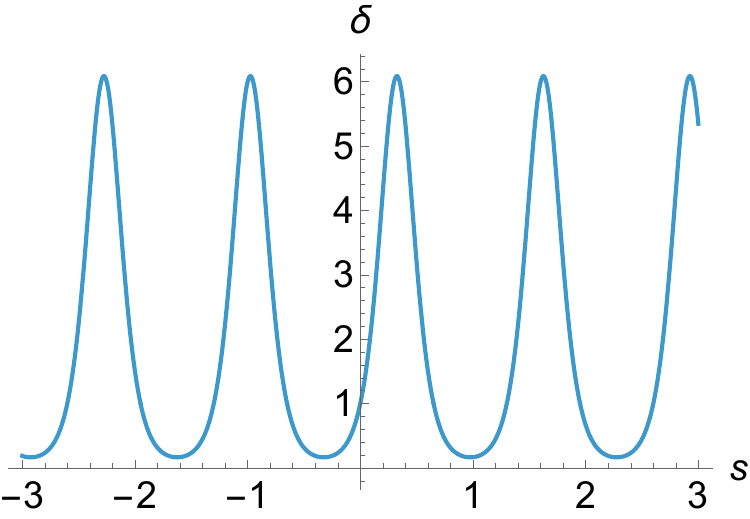}
\caption{  $\delta (s)$  for $N^2=25,c=4$ }
\end{figure}

When $N^2=25$ and $c=4$ we get the behavior for $\tau(s)$ shown in Fig. (\ref{taus}) .
For the same parameters we find that $\delta(s) =e^{\kappa} $ has the behavior shown in Fig. (\ref{deltas}) .

 Here, $s=-(b+\mu) t=-\frac{1}{2}b t  $
and $N^2=25, c=4$.

\subsection {A simple solution when the ``momentum" $\tau$  is zero}

Given the shape of the potential, there is one obvious, simple, and
surprisingly non-trivial solution, when the particle sits at the bottom
of the potential with zero momentum $\tau$.

From the equations above, the conditions for this solution are:

\bq~~\frac{d\tau}{ds} = - 4e^{\kappa} - pce^{p\kappa} = 0\eq
and \(N^{2} = 4e^{\kappa} + ce^{p\kappa}\). These two conditions
determine \(c\) and \(e^{\kappa} = \delta\). In particular, we find
from Eq. (\ref{dtaudt}) and the fact $\Omega=0$, that
\bq~~\delta = \left( \frac{- \mu}{4b} \right)N^{2} .
\eq
(Reminder: \(\mu < 0,\) so \(\delta > 0\) as it must be.)
Also, since \(\tau = 0\), we have
\bq
< \psi|\psi > =< \phi|\phi > = \frac{N}{2} . \nonumber 
\eq

From the original equations of motion, we find that

\bq~~ i\frac{d\gamma}{dt} = \left\lbrack \lambda N - (g + i\mu)\tau + \frac{1}{2}\left( \alpha_{1} - \beta_{1} \right)(N + \tau) + \frac{1}{2}\left( \alpha_{2} - \beta_{2} \right)(N - \tau) \right\rbrack\gamma ,\eq
which in our case reduces to
\bq~~ i\frac{d\gamma}{dt} = \frac{N}{2}\lbrack 2\lambda + \alpha - \beta\rbrack\gamma,
\eq
where we have abbreviated \(\alpha_{1} + \alpha_{2} = \alpha\) and
\(\beta_{1} + \beta_{2} = \beta\). Defining
\(\theta = \frac{N}{2}\lbrack 2\lambda + \alpha - \beta\rbrack\)
we have
\bq~~\gamma = \gamma_{0}e^{- i\theta t}\, \eq
with
\({\gamma_{0}}^{2} = \ \delta = \left( \frac{- \mu}{4b} \right)N^{2}\) .
(In principle, \(\gamma\) also contains an overall constant phase, which
we have here neglected.)

With \(\gamma\) in hand, we proceed to the equations of motion for the
state vectors, which in the current case become:

\bq \label{psidot} 
~~ i\frac{d|\psi >}{dt} = H|\psi > + \frac{N}{2}(\alpha + i\mu)|\psi > + (g + \lambda)\gamma|\phi >\eq

and

\bq \label{phidot} 
~~ i\frac{d|\phi >}{dt} = H|\phi > + \frac{N}{2}(\beta - i\mu)|\phi > + \left( g^{*} - \lambda \right)\gamma^{*}|\psi >.\eq

It might be thought that, once one has solved the non-linear tau-delta equations, the remaining problem of solving for the state vectors has been reduced to a linear one, 
since Eqs. (\ref{psidot}) and (\ref{phidot}) are linear. This is misleading, however, because after solving these equations, one still must impose the non-linear conditions that
$<\psi|\psi>+<\phi |\phi >=N$; $<\psi|\psi>-<\phi |\phi >=\tau $ and$<\phi | \psi >=\gamma$. 
For the case at hand, we shall do this explicitly below. 

It is convenient to expand both state vectors in eigenstates of \(H\)
(assuming a discrete spectrum for notational convenience):

\bq~~|\psi > = \sum_{n}^{}{\psi_{n}e^{- i{(E}_{n} + q)t}}|n >\eq

and

\bq~~|\phi > = \sum_{n}^{}{\phi_{n}e^{- i{(E}_{n} + q^{'})t}}|n >\eq
where \(q = \frac{N}{2}(\alpha + i\mu)\) and
\(q^{'} = \frac{N}{2}(\beta - i\mu)\) . We expect that the imaginary
parts of \(q\) and \(q^{'}\) will cancel in the final expressions for
\(|\psi >\) and \(|\phi >\).

We have

\bq~~ i{\dot{\psi}}_{n} = (g + \lambda)\gamma_{0}e^{- i\theta^{'}t}\phi_{n}\eq

and

\bq~~ i{\dot{\phi}}_{n} = \left( g^{*} - \lambda \right)\gamma_{0}e^{i\theta^{'}t}\psi_{n} ,\eq
with \(\theta^{'} = \ N(\lambda - i\mu)\).

We can decouple these equations by taking another time derivative:

\bq \label{231} 
~~{\ddot{\psi}}_{n} + iN(\lambda - i\mu){\dot{\psi}}_{n} + (g + \lambda)\left( g^{*} - \lambda \right){\gamma_{0}}^{2}\psi_{n} = 0.
\eq
This has solutions of the form \(\psi_{n} = K_{n}e^{\nu t}\), where
\(\nu\) satisfies

\bq~~\nu^{2} + i\ N(\lambda - i\mu)\nu + (g + \lambda)\left( g^{*} - \lambda \right){\gamma_{0}}^{2} = 0.
\eq
The two solutions for \(\nu\) are:
\bq~~\nu_{\pm} = - \frac{\mu N}{2} + i\chi_{\pm},\eq

where
\(\chi_{\pm} = - \frac{\lambda N}{2} \pm \frac{N}{2}\left| \sqrt{{\frac{\mu}{b}a^{2} + \mu}^{2}\left( 1 + \frac{b}{\mu} \right) - \lambda^{2}\left( 1 + \frac{\mu}{b} \right)}\  \right| \equiv - \frac{\lambda N}{2} \pm \sigma\).

With our choice of parameters, \(- b < \mu < 0\), all 3 terms under the
radical are negative. Therefore (we will need this later)

\bq~~\sigma^{2} = \frac{N^{2}}{4}\left( - \frac{\mu}{b}a^{2} - \mu^{2}\left( 1 + \frac{b}{\mu} \right) + \lambda^{2}\left( 1 + \frac{\mu}{b} \right) \right). \eq

With these inputs, we find that

\bq~~|\psi > = \sum_{n}^{}e^{- i\left( E_{n} + \frac{N}{2}\alpha + \lambda\frac{N}{2} \right)t}\left\lbrack K_{n}^{( + )}e^{i\sigma t} + K_{n}^{( - )}e^{- i\sigma t} \right\rbrack|n >\eq

and

\bq~~|\phi > = \sum_{n}^{}e^{- i\left( E_{n} + \frac{N}{2}\beta - \lambda\frac{N}{2} \right)t}\left\lbrack K_{n}^{'( + )}e^{i\sigma t} + K_{n}^{'( - )}e^{- i\sigma t} \right\rbrack|n > .
\eq

To solve the equations of motion, we must impose the following relation
between the \(K_{n}\) and the \(K_{n}^{'}\):

\bq~~ K_{n}^{'( \pm )} = \frac{i\nu_{\pm}}{(g + \lambda)\gamma_{0}}K_{n}^{( \pm )} .
\eq
Note that this proportionality factor does not depend on \(n\).

The \(K_{n}^{( \pm )}\) are not completely arbitrary, because we must
insure that

\bq~~ < \psi|\psi > = < \phi|\phi > = \frac{N}{2}\eq

and

\bq~~ < \phi|\psi > = \gamma_{0}e^{- i\theta t}   , 
\eq
where \(\ \theta = \frac{N}{2}\lbrack 2\lambda + \alpha - \beta\rbrack\).
.As we shall see, these are 4 real conditions for only 2 variables, so
it needs to be checked that they can indeed be satisfied.

In evaluating these conditions, there will be time-dependent cross terms
that can be eliminated by imposing

\bq~~ \sum_{n}^{}{K_{n}^{( - )*}K_{n}^{( + )}} = \sum_{n}^{}{K_{n}^{( + )*}K_{n}^{( - )} = 0}.
\eq
This also implies

\bq~~\sum_{n}^{}{K_{n}^{'( - )*}K_{n}^{'( + )}} = \sum_{n}^{}{K_{n}^{'( + )*}K_{n}^{'( - )} = 0}.
\eq

Define \(S_{\pm} = \sum_{n}^{}{K_{n}^{( \pm )*}K_{n}^{( \pm )}}\). Then
we have the conditions

\bq~~ \label{1eq}  < \psi|\psi > = \frac{N}{2}\ \  \Longrightarrow \ \ \ \ \ \ S_{+} + S_{-} = \frac{N}{2} , \eq

\bq~~ \label{2eq}   < \phi|\phi > = \frac{N}{2}\ \  \Longrightarrow \ \ \ \ \ \ \ \left| \nu_{+} \right|^{2}{\ S}_{+} + \left| \nu_{-} \right|^{2}\ S_{-}\ \  = \ \frac{N}{2}\ (g^{*} + \lambda)(g + \lambda){\gamma_{0}}^{2 
}, \eq

and

\bq~~   \label{3eq} < \phi|\psi > = \gamma\  \Longrightarrow \ \ \ \ \ \ \ \nu_{+}^{*}S_{+} + \nu_{-}^{*}S_{-} = i{\gamma_{0}}^{2}(g^{*} + \lambda).
\eq

These are 4 real equations for the 2 unknowns, \(S_{+}\) and \(S_{-}\).
We can dispose of one of these equations immediately. The real parts of
\(\nu_{+}^{*}\) and \(\nu_{-}^{*}\) are both \(- \frac{\mu N}{2}\), and
the real part of the right-hand side of Eq. (\ref{3eq})
 (using
\({\gamma_{0}}^{2} = \left( \frac{- \mu}{4b} \right)N^{2}\)) is
\(\left( \frac{- \mu}{4} \right)N^{2}\), so the real part of that
equation is obeyed.

The imaginary part of that equation yields an expression for
$S_{+} - S_{-}$ :

\bq~~\label{251}  S_{+} - S_{-} = \frac{N^{2}}{4\sigma}\left\lbrack \lambda\left( 1 + \frac{\mu}{b} \right) + a\frac{\mu}{b} \right\rbrack
.\eq

We can obtain another expression for \(S_{+} - S_{-}\) from Eq. (\ref{2eq}) :

\bq~~ 
S_{+} - S_{-} = \frac{1}{4\lambda\sigma}\left\lbrack \frac{N^{2}}{2}\left( \mu^{2} + \lambda^{2} \right) + 2\sigma^{2} + \frac{N^{2}}{2}\frac{\mu}{b}\left( a^{2} + b^{2} + 2\lambda a + \lambda^{2} \right) \right\rbrack .
\eq

Using the expression above for \(\sigma^{2}\) and some algebra, one can
show that these are the same, allowing for a consistent solution for
\(|\psi >\) and \(|\phi > .\)

One remaining consistency check is that, by definition, \(S_{+}\) and
\(S_{-}\) must both be positive, which requires
\(\left| S_{+} - S_{-} \right| < \frac{N}{2}\). From Eq.( {\ref{251}) , this amounts to

\bq~~\frac{4}{N^{2}}\sigma^{2} - \left\lbrack \lambda\left( 1 + \frac{\mu}{b} \right) + a\frac{\mu}{b} \right\rbrack^{2} > 0.
\eq
Working out the expression on the left-hand side using the formula for
\(\sigma^{2}\) given above, we find that it is equal to

\bq~~ - \frac{\mu}{b}\left( 1 + \frac{\mu}{b} \right)(a + \lambda)^{2} - \mu(\mu + b),
\eq
which is manifestly positive because \(- b < \mu < 0\) .

We can define two vectors that evolve according to \(H\), independently
of the non-linear terms:

\bq~~ |A > = \sum_{n}^{}{K_{n}^{( + )}e^{- iE_{n}t}}|n >\eq
and
\bq~~ |B > = \sum_{n}^{}{K_{n}^{( - )}e^{- iE_{n}t}}|n >, \eq
and then write
\bq~~ |\psi > = e^{- i\frac{N}{2}(\alpha + \lambda)t}\left\{ e^{i\sigma t}|A > + e^{- i\sigma t}|B > \right\}\eq
and
\bq~~ |\phi > = \frac{i}{(g + \lambda)\gamma_{0}}e^{- i\frac{N}{2}(\beta - \lambda)t}\left\{ \nu_{+}e^{i\sigma t}|A > + \nu_{-}e^{- i\sigma t}|B > \right\}.
\eq

\(|A >\) and \(|B >\) must obey the normalization and orthogonality
conditions

\bq~~ < A|A > + < B|B > = \frac{N}{2}\ \ \ \ \ ;\ \ \ \  < A|A > - < B|B > \  = \ \frac{N^{2}}{4\sigma}\left\lbrack \lambda\left( 1 + \frac{\mu}{b} \right) + a\frac{\mu}{b} \right\rbrack\ \eq

and \(< A|B > = 0\ .\)

These equations express the general solution for \(|\psi >\) and
\(|\phi >\), for the particular solution of the \((\tau,\delta)\)
problem that we have considered (\(\tau = 0,\ \delta = const.\)). The
fact that \(|\psi >\) is expressed more simply than \(|\phi >\) is an
artifact of the order in which we solved the equations and is not a
basic property of the system.

\section{Properties of the density matrix and the trajectory function}

In \cite{cf23}  we defined the density matrix , with its trace properly normalized to 1,  

\bq~~\ \rho\  = \frac{1}{N}\left\lbrack |\psi > < \psi| + |\phi > < \phi| \right\rbrack ,
\eq
and observed that it obeys the equation
\(i\dot{\rho} = \lbrack H,\rho\rbrack\), i.e. all the non-linear effects
disappear from \(\rho\). This raises a problem, because the most natural
extension of the Born rule to our model is to use \(\rho\) to define the
probability. However, the effects of the non-linearity would not then be
manifest in physical quantities.

Is the same true in the extended model? Using the full equations of
motion for \(|\psi >\) and \(|\phi >\) , we find

\ba~~ i\dot{\rho}&& = \lbrack H,\rho\rbrack + 2i\mu\left\lbrack \left| \psi > < \phi|\phi > < \psi \right| - |\phi > < \psi|\psi > < \phi| \right\rbrack  \nonumber\\
&&+ 2\lambda\left\lbrack \left| \phi > < \phi|\psi > < \psi \right| - |\psi > < \psi|\phi > < \phi| \right\rbrack .
\ea

In the case at hand, this becomes

\bq~~ i\dot{\rho} = \lbrack H,\rho\rbrack + i\mu N\left\lbrack |\psi > < \psi| - |\phi > < \phi| \right\rbrack + 2\lambda\left\lbrack \gamma|\phi > < \psi| - \gamma^{*}|\psi > < \phi| \right\rbrack .
\eq

We also have

\bq~~Tr\ \rho^{2} = 1 - \frac{2\mathcal{S}}{N^{2}} = 1 - \frac{c}{{2N}^{2}}\delta^{\frac{\mu}{b + \mu}} .
\eq

In conventional quantum mechanics, the deviation of the trace of $\rho^2$  from 1 is a measure of a lack of purity of the state. Here we see that this quantity is both generally less than one, and dependent on time. It is only equal to 1 if $c=0$, a limit in which $|\psi >$ and $|\phi >$   are proportional (because $\mathcal{S} =0$). 

Since \(\rho\) now includes effects of the non-linearity, it makes sense
to define the expectation value of an operator \(X\) as

\bq~~< X(t) > = Tr\ \rho(t)X = \frac{1}{N}\left\lbrack < \psi(t)|X\left| \psi(t) > + < \phi(t) \right|X|\phi(t) > \right\rbrack .
\eq

In what follows, we shall be especially interested in the case where
\(< X(t) >\) represents a trajectory, i.e. where \(X\) is the position
operator for some system. In a somewhat different context, we pursued
the idea of using trajectories of this type to investigate the classical
effects of the non-linearity in  \cite{cf24}.

To isolate the effects of the non-linearity, we want the system to move
as an otherwise free particle. We therefore take the Hamiltonian \(H\)
to be a function of \(P\) only, where \(P\) is the momentum conjugate to
\(X\). It follows that for any two vectors \(|\psi_{1} >\) and
\(|\psi_{2} >\) that evolve according to

\bq~~\left| \psi_{i}(t) > \  = e^{- iHt} \right|\psi_{i} > ,\ \ \ \ \ i = 1,2\eq

we have

\bq~~\frac{d^{2}}{dt^{2}} < \psi_{2}|X|\psi_{1} > = 0\eq

and therefore

\bq~~< \psi_{2}|X|\psi_{1} > = c_{1}t + c_{2 }.
\eq

If \({|\psi}_{1} > = |\psi_{2} >\), \(c_{1}\) and \(c_{2}\) must be real
constants. Otherwise, they can be complex.

\subsection{Trajectory function for the simple solution}

We would like to calculate the trajectory for the case of the ``simple
solution'' we have obtained above. We consider a position operator in 3
dimensions, \(X_{i}\), \(i = 1,2,3.\) The quantities
\(< A\left| X_{i} \right|A > ,\  < A\left| X_{i} \right|B >\) and
\(< B\left| X_{i} \right|B >\) will all be linear functions of time. We
are interested in the closed orbit of the particle, not in the overall
motion of the system through space (the analogy might be that we are
interested in the orbit of Earth around the sun, not the motion of the
solar system through space). Hence, we take these matrix elements to be
constants.

Using the explicit formulas for \(|\psi >\) and \(|\phi >\) , we can
express \(< \overrightarrow{X}(t) >\) as

\bq~~< {\overrightarrow{X'}}(t) > = < \overrightarrow{X}(t) > - {\overrightarrow{X}}_{0} = \overrightarrow{V}\cos{2\sigma t + \overrightarrow{W}\sin{2\sigma t}}\eq
where \({\overrightarrow{X}}_{0}\), \(\overrightarrow{V}\) and
\(\overrightarrow{W}\) are real constant vectors depending on the values we
choose for
\(< A\left| X_{i} \right|A >, \  < A\left| X_{i} \right|B >\) and
\(< B\left| X_{i} \right|B >\) . Thus, the particle moves in the plane
spanned by \(\overrightarrow{V}\) and \(\overrightarrow{W}\). Calling the
coordinates in this plane \(X_{1}^{'}\) and \(X_{2}^{'}\), we can orient
the axes so that the equation for the orbit can be written in the form

\bq~~\frac{{X_{1}^{'}}^{\mathbf{2}}}{{R_{1}}^{\mathbf{2}}}\mathbf{+}\frac{{X_{2}^{'}}^{\mathbf{2}}}{{R_{2}}^{\mathbf{2}}}\mathbf{=}1 ,\eq
i.e. the equation of an ellipse. However, this is not a Keplerian
ellipse. The particle moves harmonically about the center of the ellipse
with frequency 2\(\sigma\), and does not sweep out equal areas in equal
times as measured from one of the foci.

We expect, however, that for more general solutions to the tau-delta
problem, the trajectory motion will be less symmetrical, because the
potential itself is asymmetric (except for \(p = - 1\)).

\subsection{General formalism}

Following the same strategy as for the simple solution where \(\kappa\)
is a constant, in this section we derive the equations that will be
satisfied in the general case, although of course we do not (yet) have
solutions for these equations.

The first step is to solve the ``tau-delta'' problem, i.e. the pair of
coupled, non-linear equations

\bq~~\frac{d\kappa}{ds} = 2\tau\eq

and

\bq~~\frac{d\tau}{ds} = - 4e^{\kappa} - pce^{p\kappa} , \eq

together with the constraint

\bq~~N^{2} = h \equiv \tau^{2} + 4e^{\kappa} + ce^{p\kappa}.
\eq

Here \(s = - (b + \mu)t\) and \(h\) is the Hamiltonian:

\bq~~\frac{d\kappa}{ds} = \frac{\partial h}{\partial\tau}\eq

and

\bq~~\frac{d\tau}{ds} = - \frac{\partial h}{\partial\kappa}.\eq

The equation for \(< \phi|\psi > = \gamma\) is

\bq~~ i\frac{d\gamma}{dt} = \left\lbrack \lambda N - (g + i\mu)\tau + \frac{1}{2}\left( \alpha_{1} - \beta_{1} \right)(N + \tau) + \frac{1}{2}\left( \alpha_{2} - \beta_{2} \right)(N - \tau) \right\rbrack\gamma ,
\eq
which can be solved exactly. We define
\(\alpha_{1} + \alpha_{2} = \alpha\), \(\beta_{1} + \beta_{2} = \beta\),
\(\alpha_{1} - \alpha_{2} = \alpha',\) and
\(\beta_{1} - \beta_{2} = \beta'\). The solution is

\bq~~\gamma = e^{\frac{\kappa}{2}}e^{- i(\Lambda + \theta_{0})},\eq
where
\(\Lambda = N\left\lbrack \lambda + \frac{1}{2}(\alpha - \beta) \right\rbrack t + \frac{\kappa}{2(b + \mu)}\left\lbrack a - \frac{1}{2}(\alpha' - \beta') \right\rbrack\),
and \(\theta_{0}\) is an arbitrary real constant. We shall suppress
\(\theta_{0}\) in the following analysis.
The equations for the state vectors are
\bq~~ i\frac{d|\psi >}{dt} = \left\lbrack H + \frac{N}{2\ }((\alpha + i\mu) + \frac{\tau}{2}(\alpha' - i\mu) \right\rbrack\left| \psi > + (g + \lambda)\gamma \right|\phi >\eq
and
\bq~~ i\frac{d|\phi >}{dt} = \left\lbrack H + \frac{N}{2\ }((\beta - i\mu) + \frac{\tau}{2}(\beta' - i\mu) \right\rbrack\left| \phi > + \left( g^{*} - \lambda \right)\gamma^{*} \right|\psi > . \eq
We define
\bq~~|\psi > = e^{\frac{\mu Nt}{2} + \frac{\mu\kappa}{4(b + \mu)}}e^{- i\theta_{\psi}}|\psi^{'} > , \eq
where
\(\theta_{\psi} = \left\lbrack H + \frac{N}{2}\alpha \right\rbrack t - \frac{\kappa\alpha^{'}}{4(b + \mu)}\)
, and also
\bq~~|\phi > = e^{\frac{- \mu Nt}{2} + \frac{\mu\kappa}{4(b + \mu)}}e^{- i\theta_{\phi}}{|\phi}^{'} >, \eq
where
\(\theta_{\phi} = \left\lbrack H + \frac{N}{2}\beta \right\rbrack t - \frac{\kappa\beta^{'}}{4(b + \mu)}\).
 Then we find the coupled equations
\bq~~ i\frac{d|\psi' >}{dt} = e^{f}(g + \lambda)|\phi^{'} >\eq
and
\bq~~ i\frac{d|\phi' >}{dt} = e^{\kappa - f}\left( g^{*} - \lambda \right)|\psi^{'} >,\eq
where
\(f = \ \frac{\kappa}{2} - \mu Nt - i\left( \lambda Nt + \frac{a\kappa}{2(b + \mu)} \right) \equiv \frac{\kappa}{2} - \mu Nt - i\theta_{f}\).
Note that
\(\theta_{\phi} - \theta_{\psi} - \theta_{f} + \Lambda = 0\ .\)

We can decouple these by taking another derivative:
\bq~~\frac{d^{2}|\psi^{'} >}{dt^{2}} - \frac{df}{dt}\frac{d|\psi' >}{dt} + (g + \lambda)\left( g^{*} - \lambda \right)e^{\kappa}|\psi^{'} > = 0  . \eq
There is a similar equation for \(\frac{d^{2}|\phi^{'} >}{dt^{2}}\) , in
which \(f\) is replaced by \(f^{'} = \kappa - f\). When
\(\kappa = const.,\) we find agreement with Eq. ( \ref{231})  above.
If we assume  a solution of the form
\bq~|\psi^{'} > = F(t) |\psi_{0} > \eq
where \(|\psi_{0} >\) is a constant vector, then \(F(t)\) obeys
\bq~~\frac{d^{2}F}{dt^{2}} - \frac{df}{dt}\frac{dF}{dt} + (g + \lambda)\left( g^{*} - \lambda \right)e^{\kappa}F = 0 .\eq
This is a second-order linear differential equation, so it has 2
linearly independent solutions, \(F_{1}\) and \(F_{2}\). The general
solution for $ | \psi^{'} > $  is then

\bq~~| \psi^{'} > = F_{1}(t)  |\psi_{1} > + F_{2}(t)|\psi_{2} > ,\eq
where the \(|\psi_{i} >\) are arbitrary constant vectors. It follows
that

\bq~~|\phi' > = \frac{e^{- f}}{(g + \lambda)}i\left\lbrack \frac{dF_{1}}{dt}|\psi_{1} > + \frac{dF_{2}}{dt}|\psi_{2} > \right\rbrack .
\eq
The Wronskian of \(F_{1}\) and \(F_{2}\) has a simple form:

\bq~~W\left( F_{1},F_{2} \right) \equiv \dot{F_{1}}F_{2} - F_{1}\dot{F_{2}} = c_{0}e^{f}, \eq
where \(c_{0}\) is a constant of integration.
To recover \(|\psi >\) and \(|\phi >\) we first define
\bq~~| \psi_{i}(t) > = e^{- iHt} |\psi_{i} > \ ,\ \ i = 1,2  .\eq
These are 2 vectors that evolve according to the original Hamiltonian
\(H\). We then have

\bq~~|\ \psi > = e^{- i\frac{N}{2}(\alpha + i\mu)t}e^{\frac{i\kappa}{4}\frac{(\alpha^{'} - i\mu)}{(b + \mu)}}\left\lbrack F_{1}(t)|\psi_{1}(t) > + F_{2}(t)|\psi_{2}(t) > \right\rbrack\eq
and
\bq~~|\ \phi > = e^{- i\frac{N}{2}(\beta - i\mu)t}e^{\frac{i\kappa}{4}\frac{(\beta^{'} - i\mu)}{(b + \mu)}}\frac{e^{- f}}{(g + \lambda)}i\left\lbrack \frac{dF_{1}}{dt}|\psi_{1}(t) > + \frac{dF_{2}}{dt}|\psi_{2}(t) > \right\rbrack .\eq

If we can integrate (exactly or approximately) the tau-delta equations
and also the equations for the \(F_{i}\), the remaining steps to
achieving a full solution would then be to impose the conditions

\bq~~< \psi|\psi > + < \phi|\phi > = N\ ;\ \ \ \ \  < \psi|\psi > - < \phi|\phi > = \tau\ ;\eq

and

\bq~~< \phi|\psi > = \gamma.\eq

\section{Conclusions and outlook}

In a previous paper \cite{cf23}, we began an exploration of a particular form of
non-linear extension of quantum mechanics. The extension could be
applied to any underlying dynamical system, and featured the
introduction of two state vectors instead of the usual single state
vector in ordinary quantum mechanics. The model was simple, and the
effects of the non-linearity could be determined completely, although
the meaning of the second state vector remained uncertain. An analogy
was noted with the two-state formalism of Aharonov and Vaidman \cite{Aharonov-Vaidman} 
which is a time-symmetric formulation of ordinary quantum mechanics,
based on earlier work by Aharonov, Bergmann and Lebowitz  \cite{Aharonov64}.
We had found that our original model possessed a time-reversal symmetry, involving the interchange of 
$|\psi>$ and $ |\phi>$.  In the present extended model, this symmetry is apparently broken by the parameter $\lambda$, as we explain in the appendix.

Because of our previous model's simplicity, the dynamics it incorporated was
limited. In particular, the effects of the non-linearity persisted for
only a finite duration, decaying exponentially for large times. In
addition, the natural extension of the Born rule to the model failed to
encompass any of the non-linear behavior. Therefore, in this paper, we
have constructed an expanded version, finding the most general such
model consistent with three defining principles that were characteristic
of the earlier work. Although we are not yet able to solve this new
model completely, we can already see, by studying some simple examples,
that the time dependence of the state vectors is much richer than that
allowed by the earlier model. In particular, closed orbits exist for a
range of parameters, which was not possible when the non-linearity only
persisted for a finite time.
Previous attempts to make quantum mechanics non-linear, such as that of Weinberg, were shown to permit superluminal communication
 \cite{ Gisin} \cite{Polchinski}. That is, if two widely separated observers, Alice and Bob, were each in possession of a piece of an entangled state, Alice's measurement would instantaneously collapse the wave function describing Bob's piece. In ordinary quantum mechanics, however, Bob could not deduce from the collapse what measurement Alice made. In many non-linear extensions, though, this is no longer true, so Bob can obtain information from Alice instantaneously, violating the so-called no-signaling condition. 
Traditionally, this has been taken to be a fatal disease. The rationale given in one paper \cite{Helou} is: ``The no-signaling condition states that one cannot send information faster than the speed of light, and is a cornerstone of the special theory of relativity. The community regards this condition as being inviolable.''

We do not yet know whether our model allows this kind of superluminal communication, but if so, we do not regard it as fatal, or even necessarily detrimental. The effect will undoubtedly be hard to observe, and its discovery would reveal an interesting property of nature. We stress that physics is not determined by a vote of  ``the community'', but rather by the results of reliable experiments.

Our work so far suggests 2 approximations that may be useful in making
further progress. In this paper, we showed that, for a suitably chosen
range of parameters, the evolution of \(\kappa = \ln|\gamma|^{2}\) is
governed by a Hamiltonian whose potential,

\bq~~V(\kappa) = 4e^{\kappa} + ce^{p\kappa}\eq
rises exponentially  for $\abs{\kappa} \rightarrow \infty$.  We solved for the state vectors in the case that $\kappa$ remains stationary
at the minimum of the potential. Building on this solution, we can study small oscillations of $\kappa$ about the minimum, to see how this affects the dynamics 
of the state vectors themselves, i.e. we set
\bq~~\kappa = \kappa_{0} + \delta\kappa\eq
where \(e^{\kappa_{0}} = \left( \frac{- \mu}{4b} \right)N^{2}\) is the
minimum of the potential, and work to lowest order in \(\delta\kappa\).

Clearly, this approximation requires \(\kappa\) to remain close to the
bottom of the potential. A complementary approximation, which should be
better the farther \(\kappa\) gets from the bottom, is to replace the
true potential \(V(\kappa)\) with

\bq~~\widetilde{V}(\kappa) = 4e^{\kappa}\theta\left( \kappa - \kappa_{1} \right) + ce^{p\kappa}\theta\left( \kappa_{1} - \kappa \right)\eq

Here \(\kappa_{1}\) is defined by
\(4e^{\kappa_{1}} = ce^{p\kappa_{1}}\), so that \(\widetilde{V}\) is
continuous (this is not the same as the minimum of the potential, unless
\(p = - 1).\) The motivation for this approximation is, first, that in
each of the two regions \(\kappa > \kappa_{1}\) and
\(\kappa < \kappa_{1}\) the difference between \(V(\kappa)\) and
\(\widetilde{V}(\kappa)\) is an exponential tail that grows smaller the
larger \(|\kappa|\) becomes; and second, in each region the problem is
solvable, because the potential is given by a single exponential, which
is just the case considered in our first paper. We anticipate
implementing both these approximations in future work.

Our point of view is rather unconventional. Our first goal is to see
whether the non-linearities of our model can account for some or all of
the classical behavior of gravity. Whether or not that hope is realized,
further analysis of the extension of quantum mechanics that we have
proposed may yield valuable insight into the as yet unsolved problem of
what quantum mechanics is really all about.
\appendix
\section{Note on Time-reversal} 
Let $K$ be an anti-unitary operator that commutes with $H$ and define $K|A>=|\tilde {A} >$ for any state $|A>$. Because $ K$ is anti-unitary, we must keep track of the direction in which $K$ acts. We denote its direction 
of action with an arrow. We have $\vec{K} \vec{K}=1$, so
\bq
<B|A>=<B|\vec{K} \vec{K}|A>=<B|\overleftarrow{K}~ \overrightarrow{K}|A>^* =   < \tilde A | \tilde{B} >.
\eq

Now we define $|\psi'(t)>=K|\phi(-t)>=|\tilde{\phi} (-t)>$ and  $|\phi' (t)>=K|\psi(-t)>=|\tilde{\psi} (-t)>$.

Hence we have
\ba 
<\psi'(t)|\psi' (t)>&&=<\tilde{\phi(-t)}| \tilde{\phi} (-t)>=< \phi(-t)|\phi(-t)>  \nonumber \\
<\phi'(t)|\phi' (t)>&&=<\tilde{\psi}(-t)| \tilde{\psi} (-t)>=< \psi(-t)|\psi(-t)> ,  
\ea

and 
\bq
<\phi'(t)|\psi' (t)> =<\tilde{\psi}(-t)| \tilde{\phi} (-t)>=< \phi(-t)|\psi(-t)>
\eq
i.e. $\gamma'(t) = \gamma(-t).$

We evaluate the equations of motion at $t \rightarrow -t$ setting the $\alpha_i$ and $\beta_i$  to zero for simplicity and obtain:
\bq
- i \frac {d |\psi(-t) >}{dt} = \left[ H+ i \mu <\phi(-t)|\phi(-t)>  \right]  ~| \psi(-t) >  + (g+ \lambda ) \gamma(-t)  | \phi(-t) > 
\eq
and 
\bq
- i \frac {d |\phi(-t) >}{dt} = \left[ H- i \mu <\psi(-t)|\psi(-t)>  \right]  ~| \phi(-t) >  + (g^*-  \lambda ) \gamma^*(-t)  | \psi(-t) > .
\eq

Now we apply $K$, assuming $[K,H]=0$, and noting that $K$ will take the complex conjugate of all coefficients we then obtain" 
\bq
i \frac {d |\phi'(t) >}{dt} = \left[ H- i \mu <\psi'(t)|\psi'(t)>  \right]  ~| \phi'(t) >  + (g^*+ \lambda ) \gamma'^*(t)  | \psi'(t) > 
\eq
and 
\bq
i \frac {d |\psi'(t) >}{dt} = \left[ H+i \mu <\phi'(t)|\phi'(t)>  \right]  ~| \psi'(t) >  + (g- \lambda ) \gamma'(t)  | \phi'(t) > .
\eq

We recognize these as the equations satisfied by  $|\phi(t)> $ and $ |\psi(t)>$, respectively, except for one thing: the sign of $\lambda$  . So, modulo the problem with $\lambda$  the system possesses a time-reversal symmetry. It is not clear at this point whether there is a more general definition of the time reversal transformation that will remove the problem with $\lambda$ , or whether this is a fundamental source of time-reversal symmetry violation in our model.


\begin{thebibliography}{100} 

\bibitem{cf23} A. Chodos and F. Cooper,  Physica Scripta 98 (4), 045227 (2023).
\bibitem{Birula}  Bialynicki-Birula  and J. Mycielski, Annals of Physics 100, 62 (1976) 
\bibitem{Weinberg89} Steven Weinberg, Annals Phys., 194, 336, (1989).
\bibitem{Kaplan} D. E. Kaplan and S. Rajendran, Phys. Rev. D 105, 055002
(2022).
\bibitem{Penrose}  R. Penrose, Found. Phys. 44 , 557 (2014).
\bibitem{Carlip} S. Carlip, Class. Quantum Grav. 25, 154010  (2008).
\bibitem{Pang}  X-F Pang,   Nature Sciences  Vol. 3, No. 1, December 2008.
\bibitem{Oppenheim} J Oppenheim,   Phys. Rev.  X 13, 041040 (2023).
\bibitem{cf24} A Chodos and  F Cooper,  Symmetry  16, 887 (2024).

\bibitem{Aharonov-Vaidman} Yakir Aharonov and Lev Vaidman, 
 \bibitem{Aharonov64} Yakir Aharonov, Peter G. Bergmann, and Joel L. Lebowitz, 
Phys. Rev. 134, B1410 (1964).
Time in quantum mechanics, 399 (2008)  Springer, Berlin.
\bibitem{Gisin} N. Gisin, Phys. Lett. A 143, issues 1-2, 1(1990)
\bibitem{Polchinski} J. Polchinski, Phys. Rev. Lett. 66, 397 (1991)
\bibitem{Helou} B. Helou and Y. Chen, Journal of Physics: Conf. Series 880 (2017) 012021 
 


\end{thebibliography}
\end{document}